# SenSASP: A Unified, Multi-Layer Database of Senescence and SASP Genes

Hao Xuan[1], Yu Huang[1,2,*], Jiang Bian[1,2,3,*]

[1]Department of Biostatistics and Health Data Science, Indiana University School of Medicine, Indianapolis, IN 46202, USA

[2]Center for Biomedical Informatics, Regenstrief Institute, Indianapolis, IN 46202, USA

[3]Indiana University Health, Indianapolis, IN 46290, USA

*Correspondence: yh60@iu.edu (H.Y.), bianji@iu.edu (J.B.)

## Abstract

Research on cellular senescence and the senescence-associated secretory phenotype (SASP) draws on independently curated gene resources that differ in scope, identifiers, and update cycles, making cross-resource integration error-prone. We unified four widely used resources, CellAge, GenAge, the SenMayo signature, and the Reactome Cellular Senescence pathway, onto a single canonical identifier (the Ensembl gene ID) and enriched every gene with three annotation layers absent from all four inputs: cross-species conservation, tissue and cell-type expression, and high-confidence protein-protein interactions. Unification collapsed 1,460 summed source entries into 1,250 unique genes (210 redundant entries removed, 14.4%) while preserving full source provenance: 173 genes are corroborated by two or more resources and two (IL6, JUN) by all four. The three annotation layers reach 95.8%, 97.9%, and 93.0% of genes, with 89.4% annotated across all three. A 500-gene random sample of identifier mappings was validated against HGNC and Ensembl (98.0% exact match). The result, SenSASP, is a single, machine-readable, provenance-tracked database of harmonized identifiers and net-new functional context, illustrated here with a gene-prioritization score and a tissue-expression atlas. SenSASP is freely available at https://xuan13hao.github.io/sensasp/

## Background & Summary

Senescent cells accumulate with age and, through a pro-inflammatory secretome termed the senescence-associated secretory phenotype (SASP), contribute to tissue remodeling, tumor suppression, and a broad range of age-related pathology including neurodegenerative disease such as Alzheimer's disease [1-3]. Because eliminating or reprogramming senescent cells is being pursued as a therapeutic strategy in ageing and cancer, knowing which genes drive senescence and its secretory output is a question with direct experimental and translational consequences, not only a matter of basic cell biology. Realizing this potential requires knowing which genes to prioritize, information that is currently distributed across independently curated resources, each built for a different purpose and used by a different part of the field.

Several such resources anchor current practice. CellAge catalogs genes whose manipulation drives or prevents senescence [4]; GenAge collects genes implicated in ageing more broadly [5]; SenMayo is an experimentally derived SASP/senescence transcriptomic signature [6]; and the Reactome Cellular Senescence pathway encodes a curated mechanistic model [7]. Each reflects a genuine but different operational definition of senescence, a genetic screen, an ageing catalog, a transcriptomic signature, and a mechanistic pathway, and together these four resources are the field's primary reference points for senescence-associated genes.

These resources, however, are difficult to use together and, even individually, insufficient for translating a candidate gene into a testable hypothesis. They differ in scope, identifier conventions, and update cadence, so combining them means manually reconciling four files, resolving outdated gene symbols and duplicate entries with no shared key to join them, a task that is repeated across the field, error-prone, and rarely reported in enough detail to reproduce. More fundamentally, none of the four resources indicates whether a candidate gene is conserved in the model organisms used to test it, where in the body it is expressed, or how it is wired into the protein interaction networks that mediate SASP signaling, precisely the information a biologist needs to decide which gene, in which tissue, and in which model system to pursue next. To our knowledge, no existing resource unifies these four inputs onto a common identifier and links them to conservation, expression, and interaction data in one place.

Here we address this gap. We unify CellAge, GenAge, SenMayo, and the Reactome Cellular Senescence pathway onto a single canonical identifier (the Ensembl gene ID), removing redundant entries while recording exactly which sources support each gene, and we enrich every gene with three annotation layers absent from all four inputs: cross-species conservation, tissue and cell-type expression, and high-confidence protein–protein interactions. The result, SenSASP, is a single, machine-readable database that turns a gene-by-gene literature and database search into a direct query, which we illustrate here by prioritizing candidate senescence genes and mapping their expression across ageing-relevant tissues.

## Methods

### *Seed acquisition, Identifier harmonization, and Unification*

Four seed lists were downloaded from their primary distributors. CellAge and GenAge (human) were obtained from the Human Ageing Genomic Resources (HAGR) as versioned archives and parsed for the curated gene-symbol columns. SenMayo was retrieved as the SAUL_SEN_MAYO gene set (MSigDB accession M45803) via the MSigDB download endpoint. The Reactome Cellular Senescence pathway (R-HSA-2559583) was retrieved through the Reactome ContentService participants endpoint; gene-level participants were extracted from the reference entities of every physical entity in the pathway subtree (ReferenceDNASequence records carrying Ensembl gene IDs and ReferenceGeneProduct records carrying gene symbols).

Every unique input symbol was mapped to a canonical identifier set, current HGNC symbol [8], Ensembl gene ID, UniProt accession [9], Entrez ID [10], and HGNC ID, using MyGene.info [11] with alias-aware scopes (symbol, alias) restricted to human. When a query returned multiple hits, we preferred human records with an exact current-symbol match and an assigned Ensembl gene, breaking remaining ties by MyGene relevance score. Symbols that MyGene resolved to a different current symbol were flagged as updated.

Harmonized genes were collapsed onto the Ensembl gene ID as the canonical key (falling back to current symbol, then input symbol, for the small number of non-coding entries lacking an Ensembl mapping). For each unified gene we recorded the set of contributing sources and the set of original input symbols, yielding a per-gene source-provenance vector used for all overlap statistics.

### *Conservation, Expression, and Interaction layer*

Mouse (Mus musculus) and zebrafish (Danio rerio) orthologs were retrieved from Ensembl BioMart [12] in chunks of 200 human Ensembl gene IDs, requesting ortholog gene ID, ortholog symbol, orthology type, and percent identity. Where multiple orthologs existed, the best was selected by orthology type (one2one > one2many > many2many) then percent identity.

Human Ensembl gene IDs were resolved to versioned GENCODE IDs through the GTEx reference endpoint [13], then median TPM values were pulled across all 54 GTEx v8 tissues from the median-gene-expression endpoint. Independently, the Human Protein Atlas API supplied RNA tissue-specificity class, RNA tissue distribution, single-cell-type specificity, and subcellular location per gene [14, 15].

High-confidence protein–protein interactions were obtained from STRING v12 [16, 17] for the full set of mapped gene symbols (Homo sapiens, taxon 9606) using a combined score threshold of 700. The returned network was restricted to intra-set edges, and per-gene degree and interaction partners were computed from the resulting graph.

## Data Record

SenSASP is deposited at Zenodo (https://doi.org/10.5281/zenodo.21753677) as JSON and CSV files, comprising one record per unified gene. Each gene record contains: the canonical Ensembl ID; current HGNC symbol and gene name; an identifier block (Ensembl, UniProt, Entrez, HGNC ID, and all original input symbols); the source-provenance list and source count; and the three annotation blocks. The conservation block gives mouse and zebrafish ortholog ID, symbol, orthology type, and percent identity. The expression block gives the GENCODE ID, per-tissue GTEx median TPM across 54 tissues, and the HPA specificity and localization fields [14, 15]. The interaction block gives STRING degree, partner count, and top partners. Provenance is preserved end to end: the raw seed lists, the full identifier map, and the source-overlap matrix are all retained as intermediate artifacts, so any gene can be traced back to the exact symbols and sources it was built from.

### *Database summary statistics*

The four seed lists contributed 866 (CellAge), 307 (GenAge), 124 (SenMayo), and 163 (Reactome) genes, 1,460 entries summed across sources, spanning 1,252 unique input symbols. Identifier harmonization mapped all 1,252 symbols to Ensembl gene IDs and 1,249 to UniProt [9] accessions (the three unmapped-to-UniProt cases are non-coding genes such as TERC) and updated 5 outdated symbols to current HGNC nomenclature (e.g., ARNTL→BMAL1, CTGF→CCN2, H2AFX→H2AX). Collapsing onto the canonical key produced 1,250 unique genes, removing 210 redundant entries (14.4%) relative to the summed input.

Source provenance shows that the great majority of genes are resource-specific: 1,077 genes come from a single source, while 173 are corroborated by two or more resources, 138 by exactly two, 33 by three, and 2 (IL6 and JUN) by all four. This distribution quantifies both the fragmentation of the current landscape and the small, high-confidence core the resources agree (Figure 1).

Because none of the four foundational seed resources supply conservation, expression, or interaction metadata, all annotations across these three layers represent net-new value. Overall annotation coverage across these integrated domains is exceptionally high (Figure 2). Evolutionary conservation mapping reveals that 94.4% of genes possess a mouse ortholog and 83.6% a zebrafish ortholog, with 95.8% maintaining orthology in at least one of the two model organisms. For baseline tissue resolution, GTEx v8 median expression profiles across 54 human tissues are available for 97.9% of genes, matching the 97.9% coverage rate for Human Protein Atlas (HPA) tissue specificity annotations. Protein interaction profiling demonstrates extensive network connectivity, with 93.0% of genes participating in at least one high-confidence interaction (STRING score ≥ 700) within the set, assembling a network of 15,611 edges. The interactome topology displays marked hub-centric organization, anchored by TP53 as the top hub node (degree 368), consistent with its central regulatory role in cellular senescence shown in Figure 3. Notably, 89.4% of genes are annotated across all three functional layers simultaneously, providing a high-density, uniformly characterized core dataset well-suited for integrative systems biology analyses. Restricting to genes supported by three or more sources and clustering their GTEx profiles reveals coherent tissue-expression modules in Figure 4, including a set of broadly and highly expressed regulators alongside genes with more tissue-restricted patterns, structure that is only visible once expression is joined to the unified gene set.

## Technical Validation

Identifier-mapping accuracy was assessed with a stratified random sample of 500 of the 1,250 unified genes, drawn to give full coverage of the highest-risk and highest-interest strata (all genes supported by three or four sources, and all genes with any missing identifier field) plus a proportionally smaller random sample of two-source and one-source genes. Each sampled gene's HGNC ID, Ensembl gene ID, Entrez ID, and UniProt accession were manually cross-checked against records retrieved from the HGNC and Ensembl web portals. This audit found a 98.0% exact match rate across all identifier fields (490/500; error rate 2.0%, 95% CI 1.1–3.6%). Most discordances involved an Ensembl gene ID pointing to a duplicate gene model on an alternate haplotype contig rather than the primary assembly; the remaining discordance reflected a missing-value artifact in identifier resolution rather than an incorrect mapping.

The composite gene-prioritization score described in Usage Notes combines source confirmation (number of contributing seed resources) and network centrality (STRING degree). Because STRING degree was computed only among the curated gene set, and multiply-confirmed genes are also more likely to be well studied, these two components are moderately correlated across the full set (Spearman's $\rho = 0.33$, Pearson's $r = 0.45$, both $P < 10^{-30}$; median STRING degree rises from 12 in single-source genes to 155.5 in four-source genes). To assess whether this shared variance drives the ranking, STRING degree was residualized on source count and the ranking recomputed; 19 of the top 20 genes were unchanged, indicating the ranking is not primarily driven by this shared variance, though it should be interpreted as combining related rather than fully independent evidence.

## Usage Notes

### *Example 1: SenSASP-guided multi-evidence gene prioritization*

Integrating multiple evidence streams supports systematic candidate prioritization, a task no single source can perform alone. As one example, a composite priority score can be defined from three dimensions: source confirmation (n_sources × 25, max 100), reflecting cross-resource consensus; network centrality (STRING degree normalized to the set maximum, scaled to 50), reflecting network connectivity within the set (see Technical Validation for its relationship to source confirmation) and evolutionary conservation (mean mouse/zebrafish identity, scaled by 0.25).

Applied to all 1,250 genes, the top 20 (Figure 5) are exclusively three- or four-source genes (scores 100-142). JUN ranks first (142), combining all-four-source confirmation with strong connectivity (degree 155) and conservation (96.4% mouse identity). TP53 ranks second (141): its exceptional centrality (degree 368, the set maximum) offsets lower sequence conservation, consistent with p53's role as master regulator of stress-induced senescence [18]. IL6 ranks third (131) as the canonical SASP cytokine confirmed by all four resources [19]. The remaining top 20, CTNNB1, STAT3, EGFR, CDKN2A, ATM, MAPK14, and others, are established senescence-signaling nodes each confirmed by three sources.

This ranking is possible only because SenSASP co-locates source membership, network degree, and conservation per gene — no single seed resource, nor STRING or ortholog data alone, can reconstruct it without a unified identifier space. The score is illustrative, not prescriptive: users can reweight components, add layers (e.g., GWAS loci, single-cell co-expression), or restrict to a source subset. What SenSASP provides is turning such analyses into a query rather than a data-engineering project.

### *Example 2: aging tissue expression landscape of top-priority senescence genes*

Beyond gene prioritization, SenSASP enables direct inspection of where high-confidence senescence genes are expressed across ageing-relevant tissues. Senescence burden is known to accumulate unevenly across tissues[1, 20], and charting expression of the most evidence-supported genes across these tissues helps identify which processes are active in which contexts and which model systems are appropriate for follow-up work.

We selected 15 ageing-relevant GTEx v8 tissues, spanning brain, cardiovascular, metabolic, musculoskeletal, integumentary, renal, gastrointestinal, and haematopoietic compartments, and plotted log2(median TPM + 1) for the top 15 priority-scored genes, with rows and columns independently clustered (Figure 6).

Three modules emerge. CTNNB1, STAT3, and RELA show broad, uniform expression, consistent with constitutive regulators whose senescence-relevant activity is controlled post-translationally[21]. JUN and FOS, the AP-1 components, peak in adipose and vascular tissue, where senescence burden and AP-1-driven SASP transcription are both prominent [22]. IL6 stands apart, elevated in whole blood relative to solid tissues, reflecting its production by immune and stromal cells, a pattern invisible from network topology or source confirmation alone. The remaining low-TPM genes (ATM, CCNA2, E2F1) track cell-cycle-coupled expression, near-absent in post-mitotic tissue like brain and mature muscle.

Together, the two use cases show SenSASP's value goes beyond curation: co-locating provenance, network topology, conservation, and tissue expression turns prioritization, integrative scoring, and expression contextualization into queries rather than bespoke integration projects.

### *Limitations*

GenAge was retained in full rather than filtered to a strict senescence-only evidence subset, which would have reduced it to approximately 25 genes; the source vector is exposed so users can filter themselves. The priority-score component weights shown in Example 1 are heuristic, not empirically calibrated, and should be treated as hypothesis-generating rather than a validated ranking. The canonical key is the human Ensembl gene ID, so the small number of non-coding entries without a clean protein mapping (e.g., TERC) carry conservation and interaction annotation unevenly. Ortholog calls, expression values, and interaction scores inherit the assumptions and versions of their upstream resources (Ensembl BioMart, GTEx v8, HPA, STRING v12).

SenSASP is reproducible end to end from public APIs (see Code Availability). Natural extensions include additional model organisms, cell-type-resolved expression from single-cell atlases, and directional/regulatory edges to complement the undirected STRING network.

## Figures

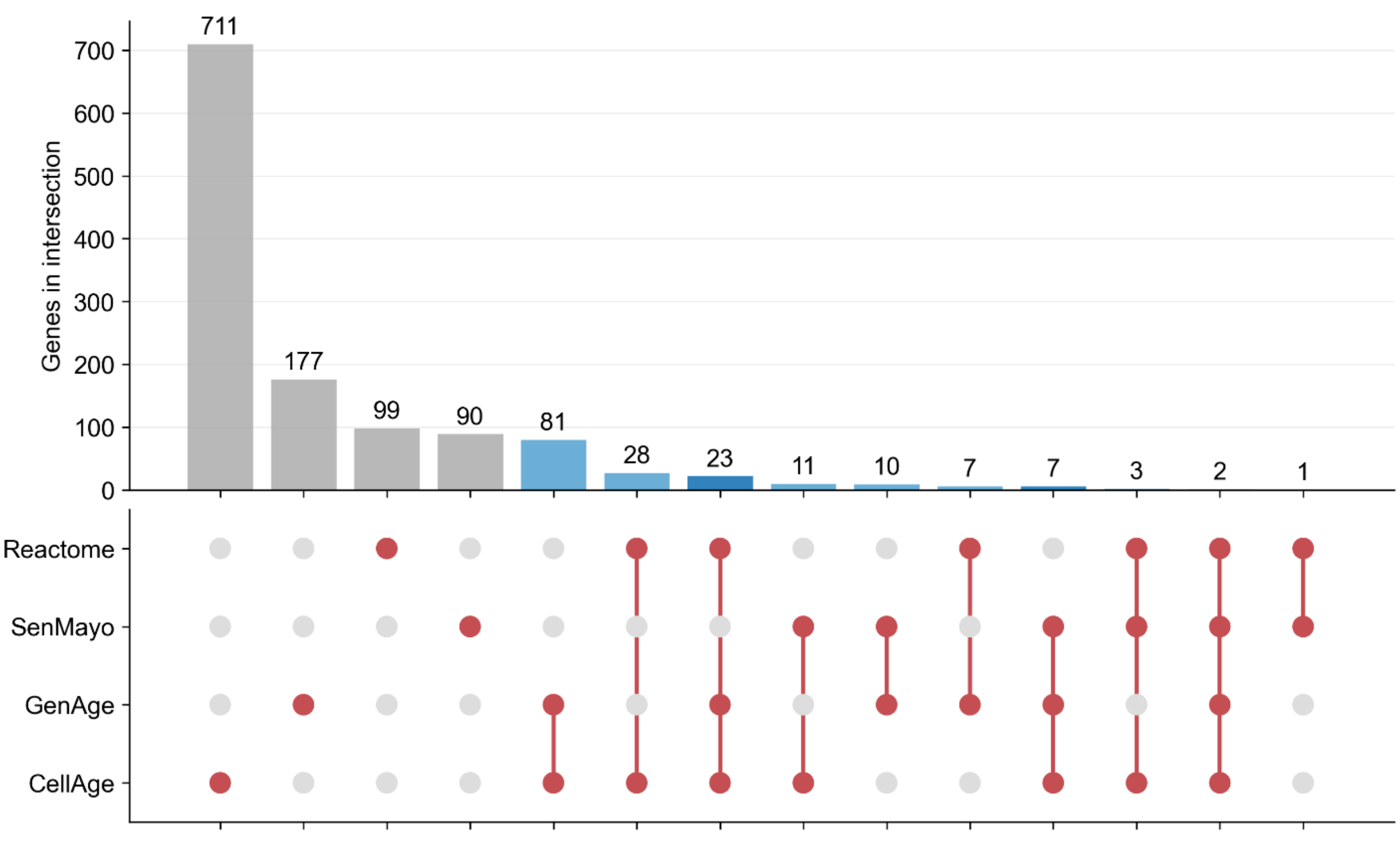

**Figure 1**: Source overlaps across the four seed lists (UpSet-style intersection sizes with source-membership matrix).

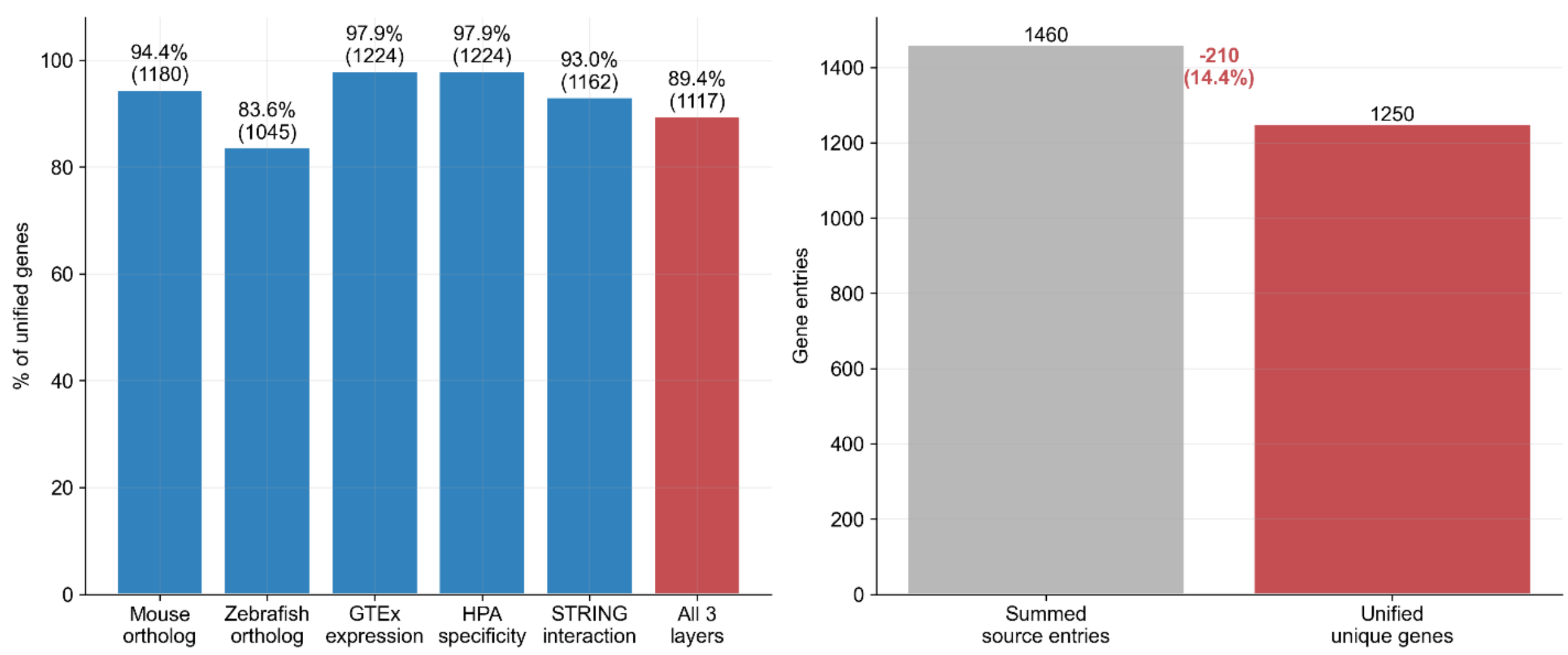

**Figure 2**: Net-new annotation coverage by layer and redundancy removed by unification.

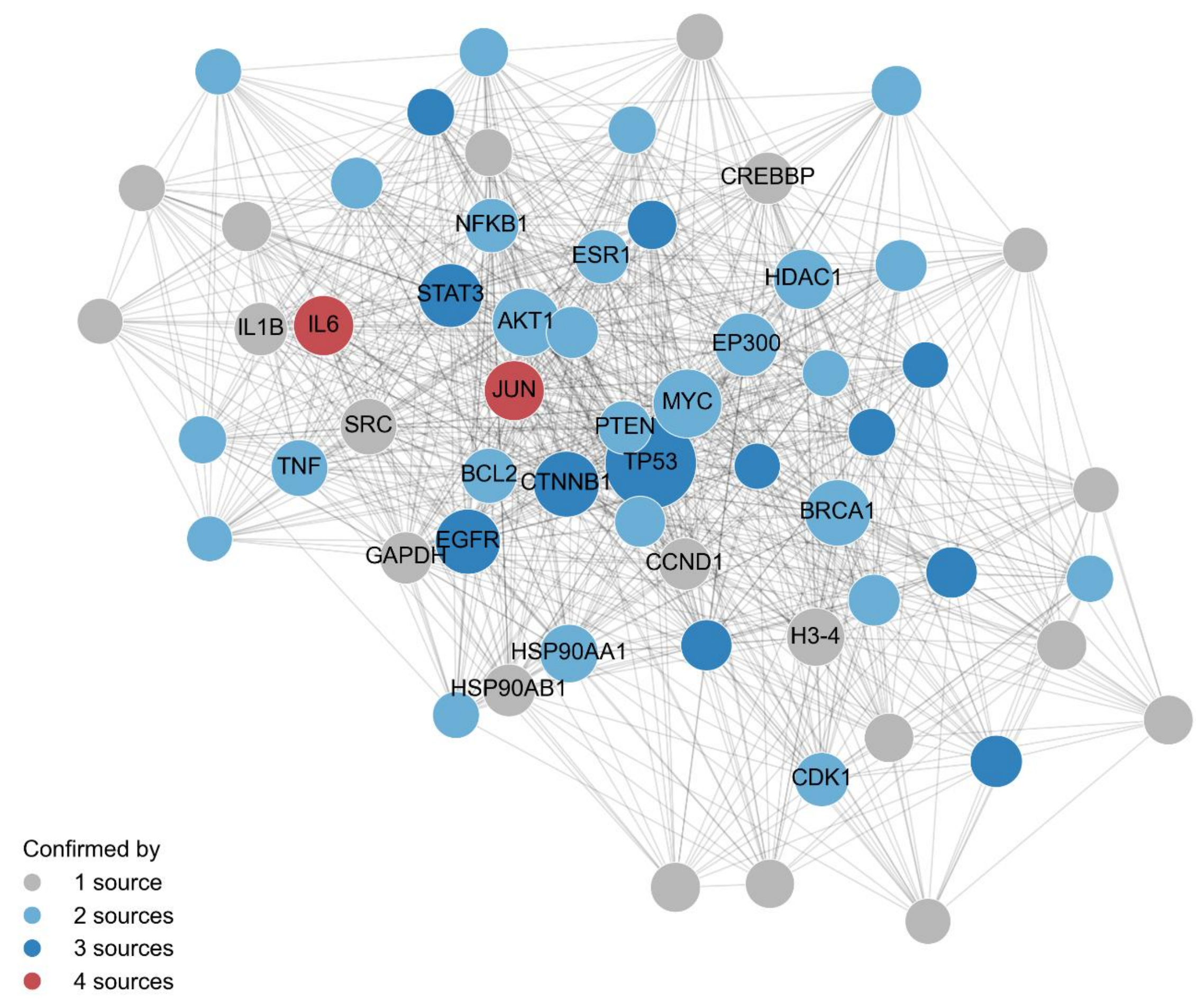

**Figure 3**: STRING v12 high-confidence interaction network for the top 60 hub genes; node size indicates degree, and node color indicates the number of confirming sources.

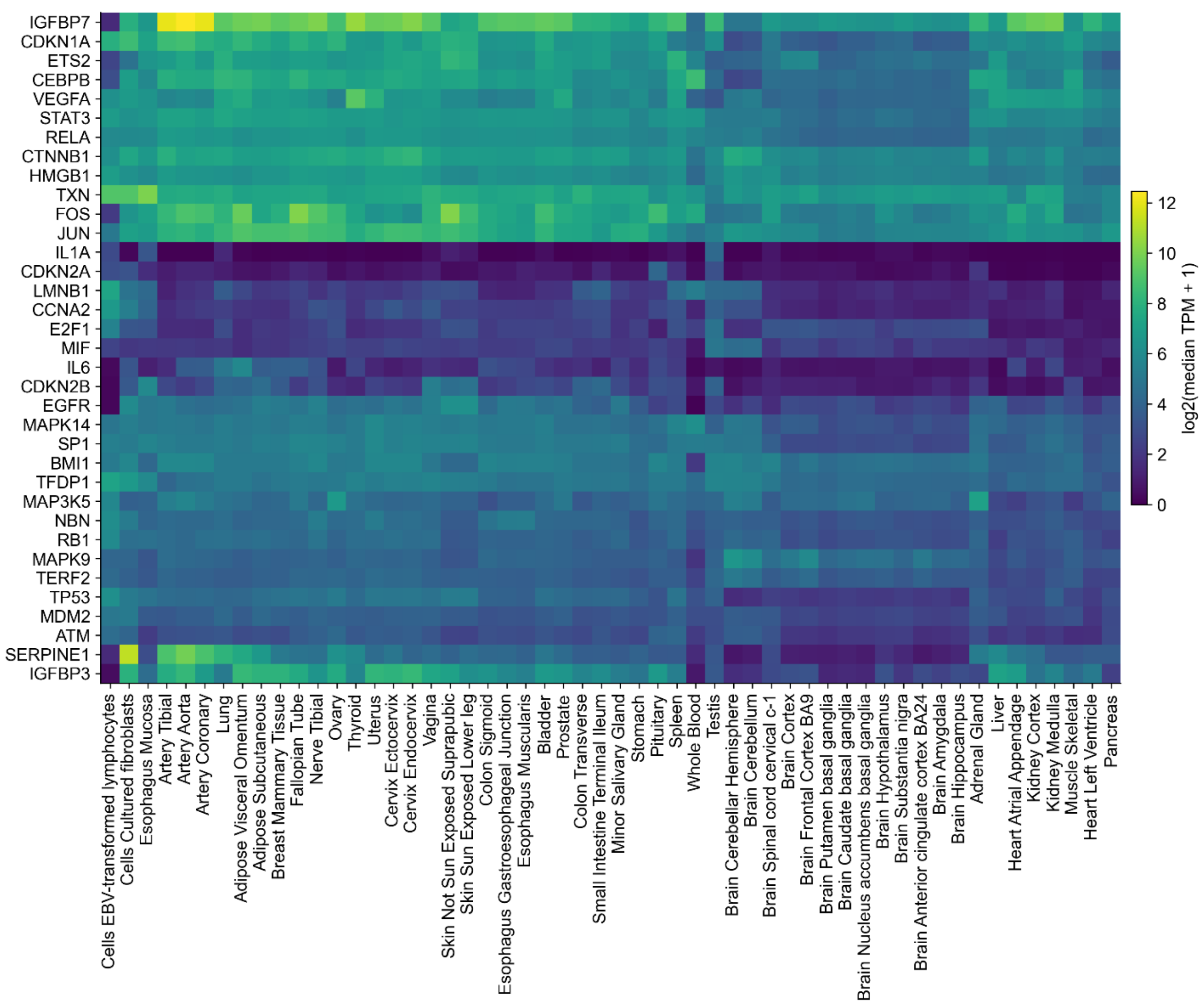

**Figure 4**: GTEx v8 median expression heatmap for genes supported by ≥3 sources, hierarchically clustered on both axes.

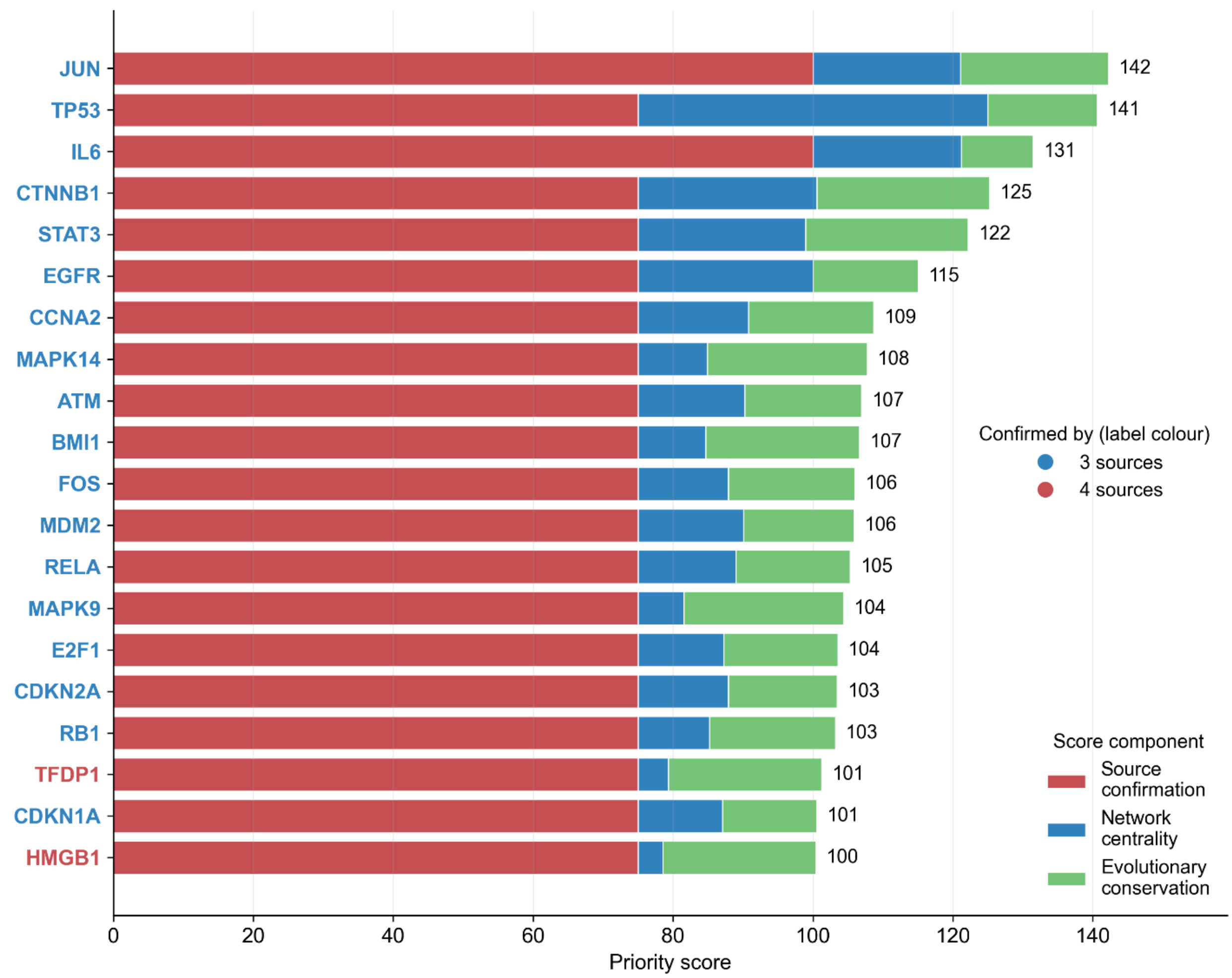

**Figure 5**: Composite priority scores for the top 20 unified genes, stacked by component (source confirmation, network centrality, evolutionary conservation); label color indicates source-confirmation tier (3 vs. 4 sources).

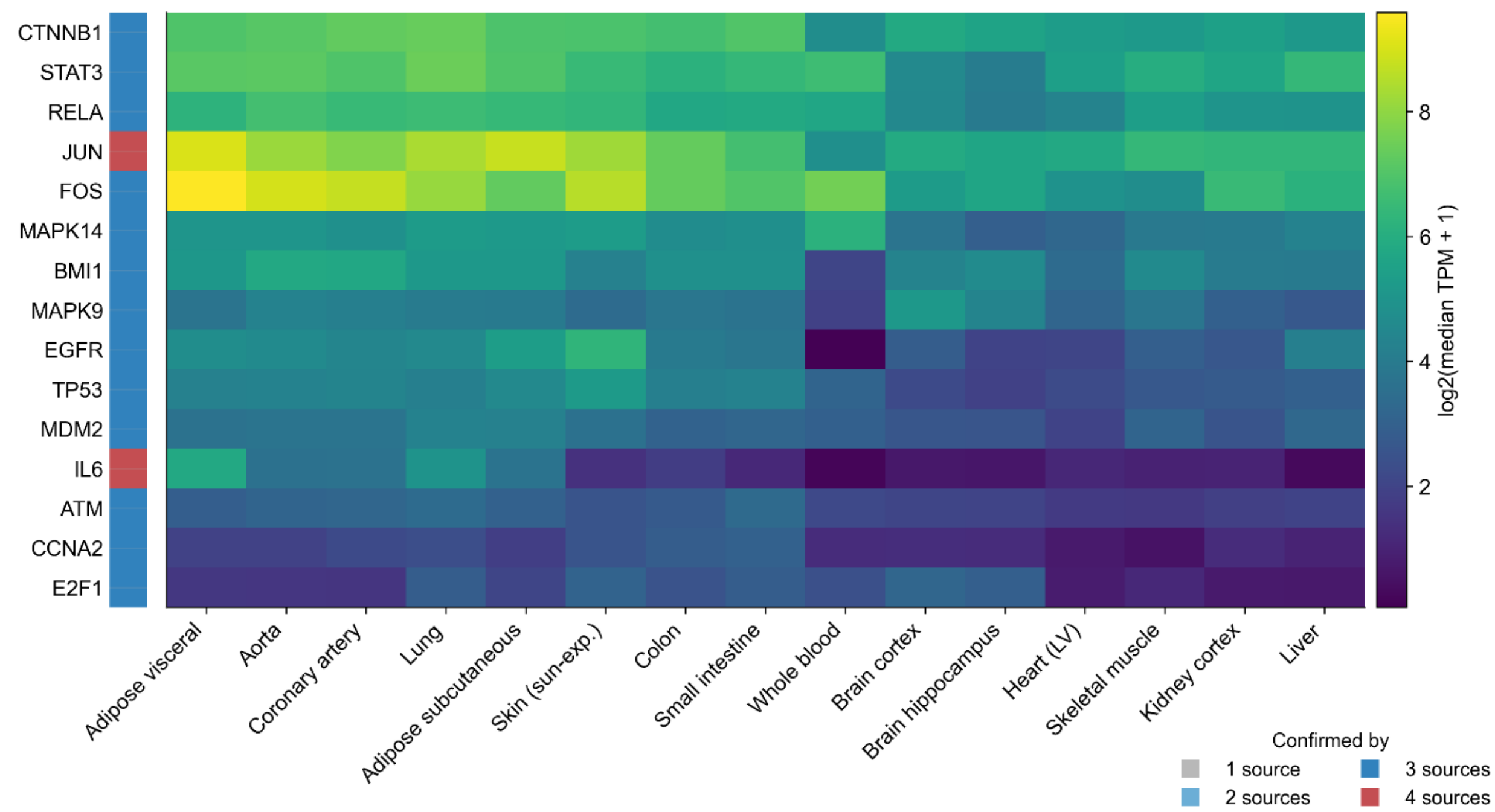

**Figure 6**: GTEx v8 median expression (log2[TPM+1]) of the top 15 priority-scored genes across 15 ageing-relevant tissues, hierarchically clustered on both axes; sidebar color indicates source-confirmation tier.

## Data availability

SenSASP, generated in this study, comprising harmonized gene records in both JSON and CSV formats, is deposited in Zenodo under a Creative Commons Attribution 4.0 International license (CC BY 4.0): https://doi.org/10.5281/zenodo.21753677. The deposited data files are senescence_sasp_database.json and senescence_sasp_database.csv, each containing one record per unified gene with the identifier, source-provenance, conservation, expression, and interaction fields described in the Data Record section above.

## Code availability

All code used to generate SenSASP and its analyses, including the full seed-acquisition, identifier-harmonization, and annotation pipeline and the scripts used to produce the figures, is deposited in the same Zenodo record as senescence_db_pipeline.zip: https://doi.org/10.5281/zenodo.21753677 (CC BY 4.0).

## Funding

This research received no specific grant from any funding agency in the public, commercial, or not-for-profit sectors.

## Competing Interests

The author declares no competing interests.

## Author Contributions

Conceptualization was carried out by HX; methodology was developed by HX. HX developed the software and generated the visualizations. Formal analysis was performed by HX; investigation was conducted by HX. Project supervision was provided by YH. The original draft was written by HX, and the final manuscript was reviewed and edited by HX, YH, and JB.

## References

1. Childs, B.G., et al., *Cellular senescence in aging and age-related disease: from mechanisms to therapy.* Nat Med, 2015. **21**(12): p. 1424-35.
2. Campisi, J., *Aging, cellular senescence, and cancer.* Annu Rev Physiol, 2013. **75**: p. 685-705.
3. Kuzniar, J., et al., *Connections Between Cellular Senescence and Alzheimer's Disease-A Narrative Review.* Int J Mol Sci, 2025. **26**(17).
4. de Magalhaes, J.P., et al., *Human Ageing Genomic Resources: updates on key databases in ageing research.* Nucleic Acids Res, 2024. **52**(D1): p. D900-D908.
5. Tacutu, R., et al., *Human Ageing Genomic Resources: integrated databases and tools for the biology and genetics of ageing.* Nucleic Acids Res, 2013. **41**(Database issue): p. D1027-33.
6. Saul, D., et al., *A new gene set identifies senescent cells and predicts senescence-associated pathways across tissues.* Nat Commun, 2022. **13**(1): p. 4827.
7. Milacic, M., et al., *The Reactome Pathway Knowledgebase 2024.* Nucleic Acids Res, 2024. **52**(D1): p. D672-D678.
8. Seal, R.L., et al., *Genenames.org: the HGNC resources in 2023.* Nucleic Acids Res, 2023. **51**(D1): p. D1003-D1009.
9. The UniProt, C., *UniProt: the universal protein knowledgebase.* Nucleic Acids Res, 2017. **45**(D1): p. D158-D169.
10. Gibney, G. and A.D. Baxevanis, *Searching NCBI databases using Entrez.* Curr Protoc Bioinformatics, 2011. **Chapter 1**: p. 1 3 1-1 3 25.
11. Wu, C., I. Macleod, and A.I. Su, *BioGPS and MyGene.info: organizing online, gene-centric information.* Nucleic Acids Res, 2013. **41**(Database issue): p. D561-5.
12. Smedley, D., et al., *BioMart--biological queries made easy.* BMC Genomics, 2009. **10**: p. 22.
13. Carithers, L.J. and H.M. Moore, *The Genotype-Tissue Expression (GTEx) Project.* Biopreserv Biobank, 2015. **13**(5): p. 307-8.
14. Uhlen, M., et al., *Towards a knowledge-based Human Protein Atlas.* Nat Biotechnol, 2010. **28**(12): p. 1248-50.
15. Ferrari, E., et al., *Age-related changes of the hypothalamic-pituitary-adrenal axis: pathophysiological correlates.* Eur J Endocrinol, 2001. **144**(4): p. 319-29.
16. Jensen, L.J., et al., *STRING 8--a global view on proteins and their functional interactions in 630 organisms.* Nucleic Acids Res, 2009. **37**(Database issue): p. D412-6.
17. Szklarczyk, D., et al., *The STRING database in 2023: protein-protein association networks and functional enrichment analyses for any sequenced genome of interest.* Nucleic Acids Res, 2023. **51**(D1): p. D638-D646.
18. Vousden, K.H. and D.P. Lane, *p53 in health and disease.* Nat Rev Mol Cell Biol, 2007. **8**(4): p. 275-83.
19. Coppe, J.P., et al., *Senescence-associated secretory phenotypes reveal cell-nonautonomous functions of oncogenic RAS and the p53 tumor suppressor.* PLoS Biol, 2008. **6**(12): p. 2853-68.
20. Lopez-Otin, C., et al., *The hallmarks of aging.* Cell, 2013. **153**(6): p. 1194-217.

21. Clevers, H. and R. Nusse, *Wnt/beta-catenin signaling and disease.* Cell, 2012. **149**(6): p. 1192-205.
22. Herranz, N. and J. Gil, *Mechanisms and functions of cellular senescence.* J Clin Invest, 2018. **128**(4): p. 1238-1246.